\documentclass[10pt]{article}
\usepackage{latexsym,graphicx}
\newcommand{\be}{\begin{equation}}
\newcommand{\ee}{\end{equation}}
\def\n{\noindent}
\catcode `\@=11
\catcode `\@=12
\begin{document}
\begin{center}
\large{\bf {Bianchi Type III String Cosmological Models with Time Dependent Bulk Viscosity}} \\
\vspace{10mm}
\normalsize{BALI Raj $^1$ and PRADHAN Anirudh$^2$}\\
\vspace{5mm}
\normalsize{$^{1}$Department of Mathematics, University of Rajasthan, Jaipur-302 004, India \\
E-mail : balir5@yahoo.co.in}\\
\vspace{5mm}
\normalsize{$^{2}$Department of Mathematics, Hindu Post-graduate College, 
Zamania-232 331, Ghazipur, India \\
E-mail : pradhan@iucaa.ernet.in, acpradhan@yahoo.com} \\
\vspace{5mm}
\vspace{5mm}
\end{center}
\vspace{10mm}
\begin{abstract} 
Bianchi type III string cosmological models with bulk viscous fluid for massive string 
are investigated. To get the determinate model of the universe, we have assumed that the 
coefficient of bulk viscosity $\xi$ is inversely proportional to the expansion $\theta$ in 
the model and expansion $\theta$ in the model is proportional to the shear $\sigma$. This 
leads to $B = \ell C^{n}$, where $\ell$ and $n$ are constants. The behaviour of the model 
in presence and absence of bulk viscosity, is discussed. The physical implications of the 
models are also discussed in detail. 
\end{abstract}
 \smallskip
\n Keywords: Cosmic string, viscous models, Bianchi type III model\\
\n PACS: 98.80.Cq, 04.20.-q 
\section{Introduction}
In recent years, there has been considerable interest in string cosmology. Cosmic 
strings are topologically stable objects which might be found during a phase 
transition in the early universe (Kibble [1]). Cosmic strings play an important role in 
the study of the early universe. These arise during the phase transition after the 
big bang explosion as the temperature goes down below some critical temperature as 
predicted by grand unified theories (Zel'dovich et al. [2]; Kibble [1, 3]; 
Everett [4]; Vilenkin [5]). It is believed that cosmic strings give rise to density 
perturbations which lead to the formation of galaxies (Zel'dovich [6]). These cosmic 
strings have stress-energy and couple to the gravitational field. Therefore it is interesting to 
study the gravitational effects that arise from strings. 

The general relativistic treatment of strings was initiated by Letelier [7, 8] and 
Stachel [9]. Letelier [7] has obtained the solution to Einstein's field equations 
for a cloud of strings with spherical, plane and cylindrical symmetry. Then, in 1983, he 
solved Einstein's field equations for a cloud of massive strings and obtained cosmological 
models in Bianchi I and Kantowski-Sachs space-times. Benerjee et al. [10] have investigated 
an axially symmetric Bianchi type I string dust cosmological model in presence and absence 
of magnetic field. The string cosmological models with a magnetic field are also discussed 
by Chakraborty [11], Tikekar and Patel [12]. Patel and Maharaj [13] investigated 
stationary rotating world model with magnetic field. Ram and Singh [14] obtained some new exact 
solutions of string cosmology with and without a source free magnetic field for Bianchi type I 
space-time in the different basic form considered by Carminati and McIntosh [15]. Exact 
solutions of string cosmology for Bianchi type II, $VI_0$, VIII and IX space-times have been 
studied by Krori et al. [16] and Wang [17]. Singh and Singh [18] investigated string 
cosmological models with magnetic field in the context of space-time with $G_{3}$ 
symmetry. Lidsey, Wands and Copeland [19] have reviewed aspects of super string 
cosmology with the emphasis on the cosmological implications of duality symmetries in 
the theory. Baysal et al. [20] have investigated the behaviour of a string in the 
cylindrically symmetric inhomogeneous universe. Bali et al. [21 - 25]  have 
obtained Bianchi types IX, type-V and type-I string cosmological models in general relativity. Yavuz 
[26] have examined charged strange quark matter attached to the string cloud in the 
spherical symmetric space-time admitting one-parameter group of conformal motion. Recently 
Kaluza-Klein cosmological solutions are obtained by Yilmaz [27] for quark matter coupled 
to the string cloud in the context of general relativity.

On the other hand, the matter distribution is satisfactorily described by perfect fluids due 
to the large scale distribution of galaxies in our universe. However, a realistic treatment 
of the problem requires the consideration of material distribution other than the perfect 
fluid. It is well known that when neutrino decoupling occurs, the matter behaves as a viscous 
fluid in an early stage of the universe. Viscous fluid cosmological models of early universe 
have been widely discussed in the literature.   

Recently, Seljak et al. [28] have studied cosmological parameters from combining Lyman-$\alpha$ 
forest with C.M.B. They established the cosmic strings limited to 
$ G\mu < 2.3 \times 10^{-7} $ at  95 percent  cathodoluminescence (c.l.),  and correlated iso-curvature 
models are also tightly constrained. Observations on C.M.B. constrain the cosmic string which is 
predicted in terms of cosmic string power spectrum [29]. Howerever, we are not concerned 
with this new phenomena. Tikekar and Patel [12], following the techniques used by Letelier and Stachel, 
obtained some exact Bianchi III cosmological solutions of massive strings in presence of magnetic field. 
Maharaj et al. [30] have generalized the previous solutions obtained by Tikekar and Patel [12]  
considering Lie point symmetries. Yadav et al. [31] have studied some Bianchi type I viscous fluid 
string cosmological models with magnetic field. Recently Wang [32 - 35] has also discussed LRS Bianchi 
type I and Bianchi type III cosmological models for a cloud string with bulk viscosity. Very recently, 
Yadav, Rai and Pradhan [36] have found the integrability of cosmic string in Bianchi type III space-time 
in presence of bulk viscous fluid by applying a new technique. Motivated by the situations 
discussed above, in this study, we established a formalism for studying the new 
integrability of massive strings in Bianchi III space-time in presence of a variable bulk 
viscosity. 
\section{The Metric and Field  Equations}
We consider the space-time of general Bianchi III type with the metric 
\begin{equation}
\label{eq1}
ds^{2} =  - dt^{2} + A^{2}(t) dx^{2} + B^{2}(t) e^{-2 \alpha x} dy^{2} + C^{2}(t) dz^{2},
\end{equation}
where $\alpha$ is constant. The energy momentum tensor for a cloud of string dust 
with a bulk viscous fluid of string is given by Letelier [7] and Landau \& 
Lifshitz [37] 
\begin{equation}
\label{eq2}
T^{j}_{i} = \rho v_{i}v^{j} - \lambda x_{i}x^{j} - \xi v^{l}_{;l} \left(g^{j}_{i} 
+ v_{i} v^{j}\right),
\end{equation}
where $v_{i}$ and $x_{i}$ satisfy condition
\begin{equation}
\label{eq3}
v^{i} v_{i} = - x^{i} x_{i} = -1, \, \, \, v^{i} x_{i} = 0,
\end{equation}
where $\rho$ is the proper energy density for a cloud string with 
particles attached to them, $\lambda$ is the string tension density, $v^{i}$ is the 
four-velocity of the particles, and $x^{i}$ is a unit space-like vector representing 
the direction of string. If the particle density of the configuration is denoted by 
$\rho_{p}$, then we have
\begin{equation}
\label{eq4}
\rho = \rho_{p} + \lambda.
\end{equation}
The Einstein's field equations (in gravitational units $c = 1$, $G = 1$) read 
\begin{equation}
\label{eq5}
R^{j}_{i} - \frac{1}{2} R g^{j}_{i} = - 8\pi T^{j}_{i},
\end{equation}
where $R^{j}_{i}$ is the Ricci tensor; $R$ = $g^{ij} R_{ij}$ is the
Ricci scalar. In a co-moving co-ordinate system, we have
\begin{equation}
\label{eq6}
v^{i} = (0, 0, 0, 1), \, \, \, x^{i} = (0, 0, 1/C, 0).
\end{equation}
The field equations (\ref{eq5}) with Eq. (\ref{eq2}) subsequently lead to the following 
system of equations:
\begin{equation}
\label{eq7}
\frac{B_{44}}{B} + \frac{C_{44}}{C} + \frac{B_{4} C_{4}}{B C} = 8\pi \xi \theta,
\end{equation}
\begin{equation}
\label{eq8}
\frac{A_{44}}{A} + \frac{C_{44}}{C} + \frac{A_{4} C_{4}}{A C} = 8\pi \xi \theta,
\end{equation}
\begin{equation}
\label{eq9}
\frac{A_{44}}{A} + \frac{B_{44}}{B} + \frac{A_{4} B_{4}}{A B} - \frac{a^{2}}{A^{2}} = 
8\pi (\lambda + \xi \theta),
\end{equation}
\begin{equation}
\label{eq10}
\frac{A_{4}B_{4}}{A B} + \frac{B_{4} C_{4}}{B C} + \frac{C_{4} A_{4}}{C A} - \frac{a^{2}}
{A^{2}}  = 8\pi \rho,
\end{equation}
\begin{equation}
\label{eq11}
\frac{A_{4}}{A} - \frac{B_{4}}{B} = 0,
\end{equation}
where the subscript $4$ at the symbols $A$, $B$ and $C$ denotes ordinary differentiation 
with respect to $t$.
The particle density $\rho_{p}$ is given by 
\begin{equation}
\label{eq12}
8\pi \rho_{p} = \frac{B_{4} C_{4}}{B C} + \frac{C_{4} A_{4}}{C A} - \frac{B_{44}}{B} 
- \frac{A_{44}}{A} + 8\pi K
\end{equation}
in accordance with Eq. (\ref{eq4}). \\

The velocity field $v^{i}$ as specified by Eq. (\ref{eq6}) is irrotational. The 
expansion ($\theta$) and components of shear tensor ($\sigma_{ij}$) are given by
\begin{equation}
\label{eq13}
\theta = \frac{A_{4}}{A} + \frac{B_{4}}{B} + \frac{C_{4}}{C},
\end{equation}
\begin{equation}
\label{eq14}
\sigma^{1}~~_{1} = \frac{1}{3}\left[\frac{2A_{4}}{A} - \frac{B_{4}}{B} - \frac{C_{4}}{C}\right],
\end{equation}
\begin{equation}
\label{eq15}
\sigma^{2}~~_{2} = \frac{1}{3}\left[\frac{2B_{4}}{B} - \frac{A_{4}}{A} - \frac{C_{4}}{C}\right],
\end{equation}
\begin{equation}
\label{eq16}
\sigma^{3}~~_{3} = \frac{1}{3}\left[\frac{2C_{4}}{C} - \frac{A_{4}}{A} - \frac{B_{4}}{B}\right],
\end{equation}
\begin{equation}
\label{eq17}
\sigma^{4}~~_{4} = 0.
\end{equation}
Therefore
$$\sigma^{2} = \frac{1}{2}\left[(\sigma^{1}~~_{1})^{2} + (\sigma^{2}~~_{2})^{2} + 
(\sigma^{3}~~_{3})^{2} + (\sigma^{4}~~_{4})^{2}\right]$$
leads to
\begin{equation}
\label{eq18}
\sigma^{2} = \frac{1}{3}\left[\frac{A_{4}^{2}}{A^{2}} + \frac{B_{4}^{2}}{B^{2}} + 
\frac{C_{4}^{2}}{C^{2}} - \frac{A_{4}B_{4}}{AB} - \frac{B_{4}C_{4}}{BC} - 
\frac{A_{4}C_{4}}{AC}\right]. 
\end{equation}
\section{Solutions of the Field  Equations}
The field equations (\ref{eq7})-(\ref{eq11}) are a system of five equations with six 
unknown parameters $A$, $B$, $C$, $\rho$, $\lambda$ and $\xi$. One additional constraint 
relating these parameters is required to obtain explicit solutions of the system. Referring 
to Thorn [38], observations of the velocity-redshift relation for extragalactic sources suggest 
that Hubble expansion of the universe is isotropic today to within about $30$ percent [39, 40]. 
Put more precisely, redshift studies place the limit 
$$\frac{\sigma}{H} \leq 0.30$$ 
on the ratio of shear, $\sigma$, to Hubble constant $H$ in the neighbourhood of our Galaxy. Following 
Bali and Jain [41] and Pradhan et al. [42, 43], we assume that the expansion ($\theta$) in the model is 
proportional to the shear ($\sigma$), which is physically plausible condition. This condition leads to 
\begin{equation}
\label{eq19}
B = \ell C^{n}
\end{equation}
where $\ell$ is proportionality constant and $n$ is a constant. Equations (\ref{eq11}) leads to
\begin{equation}
\label{eq20}
A = m B,
\end{equation}
where $m$ is an integrating constant. To obtain the determinate model of the universe, we assume 
that the coefficient of bulk viscosity is inversely proportional to expansion ($\theta$). This 
condition leads to 
\begin{equation}
\label{eq21}
\xi \theta = K, 
\end{equation}
where $K$ is a proportionality constant.

By the use of Eqs. (\ref{eq19}) and (\ref{eq21}), Eq. (\ref{eq7}) leads to 
\begin{equation}
\label{eq22}
2C_{44} + \left(\frac{2n^{2}}{n + 1}\right)\frac{C_{4}^{2}}{C} = \left(\frac{16 \pi K}
{n + 1}\right) C.
\end{equation}
Let
\begin{equation}
\label{eq23}
C_{4} = f(C). 
\end{equation}
Hence Eq. (\ref{eq22}) leads to
\begin{equation}
\label{eq24}
\frac{d}{dC}\left(f^{2}\right) + \left(\frac{2n^{2}}{n + 1}\right)\frac{f^{2}}{C} =  
\left(\frac{16 \pi K}{n + 1}\right) C.
\end{equation}
From Eq. (\ref{eq24}), we have
\begin{equation}
\label{eq25}
f^{2}C^{\frac{2n^{2}}{n + 1}} = \left(\frac{8 \pi K}{n^{2} + n + 1}\right) 
C^{\frac{2n^{2} + 2n + 2}{n + 1}} + L,
\end{equation}
where $L$ is an integrating constant. Eq. (\ref{eq25}) leads to
\begin{equation}
\label{eq26}
f^{2} = \left(\frac{8\pi K}{n^{2} + n + 1}\right)C^{2} + L C^{-(\frac{2n^{2}}{n + 1})}.
\end{equation}
From Eq. (\ref{eq26}), we have 
\begin{equation}
\label{eq27}
\frac{C^{\frac{n^{2}}{n + 1}}dC}{\sqrt{C^{2\left(\frac{n^{2} + n + 1}{n + 1}\right)} + b^{2}}} 
= \sqrt{a} dt.
\end{equation}
Integrating Eq. (\ref{eq27}), we have
\begin{equation}
\label{eq28}
C = \left[\frac{\beta^{2}(\beta t + \gamma)^{2} - b^{2}}{\beta^{2}}\right]^{\left(\frac{n + 1}
{n^{2} + n + 1}\right)},
\end{equation}
where $\gamma$ is an integrating constant and 
$$ \beta = \frac{(n^{2} + n + 1)\sqrt{a}}{2(n + 1)},$$
$$ a = \frac{8\pi K}{(n^{2} + n + 1))},$$
$$ b^{2} = \frac{L}{a}.$$
Therefore, we have
\begin{equation}
\label{eq29}
B = \ell C^{n} = \ell \left[\frac{\beta^{2}(\beta t + \gamma)^{2} - b^{2}}{\beta^{2}}\right]^
{\left(\frac{n(n + 1)}{n^{2} + n + 1}\right)},
\end{equation}
\begin{equation}
\label{eq30}
A = m B = \ell m  \left[\frac{\beta^{2}(\beta t + \gamma)^{2} - b^{2}}{\beta^{2}}\right]^
{\left(\frac{n(n + 1)}{n^{2} + n + 1}\right)}.
\end{equation}
Hence the metric (\ref{eq1}) reduces to the form
\[
ds^{2} = - dt^{2} + \left[\frac{\beta^{2}(\beta t + \gamma)^{2} - b^{2}}{\beta^{2}}\right]
^{\frac{2n(n + 1)}{n^{2} + n + 1}}\left[ dX^{2} + e^{-2\alpha X} dY^{2}\right]
\]
\begin{equation}
\label{eq31}
+ \left[\frac{\beta^{2}(\beta t + \gamma)^{2} - b^{2}}{\beta^{2}} \right]^{\frac{2(n + 1)}
{n^{2} + n + 1}}dZ^{2},
\end{equation}
where
$$\ell m x = X, ~ ~ \ell y = Y, ~ ~ z = Z. $$
Using the transformation
\begin{equation}
\label{eq32}
\beta(\beta t + \gamma) = b \cos{(\beta \tau)}
\end{equation}
the metric (\ref{eq31}) reduces to the form
\[
ds^{2} = - \frac{b^{2}\sin^{2}{(\beta \tau)}}{\beta^{2}} d\tau^{2} + \left[\frac{b^{4}\sin^{4}
{(\beta \tau)}}{\beta^{4}}\right]^{\frac{n(n + 1)}{n^{2} + n + 1}}\left[dX^{2} + e^{-\frac{2\alpha X}
{\ell m}} dY^{2}\right]
\]
\begin{equation}
\label{eq33}
+ \left[\frac{b^{4}\sin^{4}{(\beta \tau)}}{\beta^{4}}\right]^{\frac{(n + 1)}{n^{2} + n + 1}} dZ^{2}.
\end{equation}
In absence of viscosity i.e. when $\beta \to 0$ then the metric (\ref{eq33}) reduces to the new form
\begin{equation}
\label{eq34}
ds^{2} = -b^{2}\tau^{2} d\tau^{2} + \left(b^{4} \tau^{4}\right)^{\frac{n(n + 1)}{n^{2} + n + 1}}
\left[dX^{2} + e^{-\frac{2\alpha X}{\ell m}} dY^{2}\right]+ \left(b^{4} \tau^{4}\right)^
{\frac{(n + 1)}{n^{2} + n + 1}} dZ^{2}.
\end{equation}
\section{Some Physical and Geometrical Features of the Models}
The rest energy density ($\rho$), the string tension density ($\lambda$), the particle density 
($\rho_{p}$), the scalar of expansion ($\theta$) and the shear ($\sigma$) for the model (\ref{eq31}) 
are given by
\begin{equation}
\label{eq35}
8\pi \rho = \frac{4n(n + 2)(n + 1)^{2}}{(n^{2} + n + 1)^{2}} \frac{\beta^{2}(\beta t + \gamma)^{2}}
{\left[\frac{\beta^{2}(\beta t + \gamma)^{2} - b^{2}}{\beta^{2}}\right]^{2}} - \frac{\alpha^{2}}
{m^{2}\ell^{2}\left[\frac{\beta^{2}(\beta t + \gamma)^{2} - b^{2}}{\beta^{2}}\right]^
{\frac{2n (n + 1)}{n^{2} + n + 1}}}, 
\end{equation}
\[
8\pi \lambda = \frac{4n(n + 1)\beta^{2}}{(n^{2} + n + 1)\left[\frac{\beta^{2}(\beta t + \gamma)^{2}
 - b^{2}}{\beta^{2}}\right]} + \frac{4n(n + 1)\beta^{2}(\beta t + \gamma)^{2} (n^{2} + n - 2)}
{(n^{2} + n + 1)^{2}\left[\frac{\beta^{2}(\beta t + \gamma)^{2} - b^{2}}{\beta^{2}}\right]^{2}}
\]
\begin{equation}
\label{eq36}
- \frac{\alpha^{2}}{m^{2}\ell^{2}\left[\frac{\beta^{2}(\beta t + \gamma)^{2} - b^{2}}{\beta^{2}}
\right]^{\frac{2n (n + 1)}{n^{2} + n + 1}}} - \frac{\beta^{2}(n + 1)^{2}}{(n^{2} + n +1)}, 
\end{equation}
\[
8\pi \rho_{p} = \frac{8n(n + 1)(n + 2)\beta^{2}(\beta t + \gamma)^{2}}{(n^{2} + n + 1)^{2}
\left[\frac{\beta^{2}(\beta t + \gamma)^{2} - b^{2}}{\beta^{2}}\right]^{2}}
\]
\begin{equation}
\label{eq37}
 - \frac{4n(n + 1)\beta^{2}}{(n^{2} + n + 1)\left[\frac{\beta^{2}(\beta t + \gamma)^{2}
 - b^{2}}{\beta^{2}}\right]} + \frac{\beta^{2}(n + 1)^{2}}{(n^{2} + n +1)}, 
\end{equation}
\begin{equation}
\label{eq38}
\theta = \frac{2(2n + 1)(n + 1)\beta (\beta t + \gamma)}{(n^{2} + n + 1)
\left[\frac{\beta^{2}(\beta t + \gamma)^{2}- b^{2}}{\beta^{2}}\right]},
\end{equation}
\begin{equation}
\label{eq39}
\sigma  = \frac{2(n^{2} - 1)\beta (\beta t + \gamma)}{(n^{2} + n + 1)\sqrt{3}
\left[\frac{\beta^{2}(\beta t + \gamma)^{2}- b^{2}}{\beta^{2}}\right]}.
\end{equation}

The energy conditions $\rho \geq 0$ and $\rho_{p} \geq 0$ are satisfied in the presence of 
bulk viscosity for the model (\ref{eq31}). The condition $\rho \geq 0$ leads to
$$- \frac{\alpha (n^{2} + n +1)}{2\ell m (n + 1)\sqrt{n(n + 2)}} \leq \frac{\beta(\beta t + \gamma)}
{\left[\frac{\beta^{2}(\beta t + \gamma)^{2} - b^{2}}{\beta^{2}}\right]^{\frac{1}{n^{2} + n + 1}}} 
\leq \frac{\alpha (n^{2} + n +1)}{2\ell m (n + 1)\sqrt{n(n + 2)}}.$$ 
From Eq. (\ref{eq36}), we observe that the string tension density $\lambda \geq 0$ provided
$$\frac{1}{\left[\frac{\beta^{2}(\beta t + \gamma)^{2} - b^{2}}{\beta^{2}}\right]} + 
\frac{(\beta t + \gamma)^{2}(n^{2} + n - 2)}{(n^{2} + n + 1)\left[\frac{\beta^{2}(\beta t + \gamma)
^{2} - b^{2}}{\beta^{2}}\right]} \geq $$
$$\frac{\alpha^{2} (n^{2} + n + 1)}{4n(n + 1) m^{2} \ell^{2}\beta^{2}{\left[\frac{\beta^{2}
(\beta t + \gamma)^{2} - b^{2}}{\beta^{2}}\right]^{\frac{2n(n + 1)}{n^{2} + n + 1}}}} + 
\frac{(n + 1)}{4n}. $$
The model in presence of bulk viscosity starts with a big bang at time $t = \frac{b}{\beta^{2}} 
- \frac{\gamma}{\beta}$. The expansion in the model decreases as time increases. The expansion 
in the model stops at $t = -\frac{\gamma}{\beta}$. The model (\ref{eq31}) in general represents 
shearing and non-rotating universe. The role of bulk viscosity is to retard expansion in the model.  
We can see from the above discussion that the bulk viscosity plays a significant role in the 
evolution of the universe. Furthermore, since $\lim_{t \to \infty} \frac{\sigma}{\theta} \ne 0$, 
the model does not approach isotropy for large value of $t$. However the model isotropizes when 
$t = -\frac{\gamma}{\beta}$. There is a real physical singularity in the model (\ref{eq31}) at 
$t = 0$.  

Using the transformation
$$\beta(\beta t + \gamma) = b \cos{(\beta \tau)},$$
the rest energy density ($\rho$), the string tension density ($\lambda$), the particle density 
($\rho_{p}$), the scalar of expansion ($\theta$) and the shear ($\sigma$) for the model (\ref{eq33}) 
are given by
\begin{equation}
\label{eq40}
8\pi \rho = \frac{4n(n + 2)(n + 1)^{2}\cos^{2}{(\beta \tau)}}{b^{2}(n^{2} + n + 1)^{2}
\left[\frac{\sin^{4}{(\beta \tau)}}{\beta^{4}}\right]} - \frac{\alpha^{2}}{m^{2}\ell^{2}\left[b^{2}
\frac{\sin^{2}{(\beta \tau)}}{\beta^{2}}\right]^{\frac{2n (n + 1)}{n^{2} + n + 1}}}, 
\end{equation}
\[
8\pi \lambda = - \frac{4n(n + 1)\beta^{2}}{(n^{2} + n + 1)\left[b^{2}\frac{\sin^{2}{(\beta \tau)}}
{\beta^{2}}\right]} + \frac{4n(n + 1)(n^{2} + n - 2)b^{2}\cos^{2}{(\beta \tau)}}
{(n^{2} + n + 1)^{2}\left[b^{2}\frac{\sin^{2}{(\beta \tau)}}{\beta^{2}}\right]^{2}}
\]
\begin{equation}
\label{eq41}
- \frac{\alpha^{2}}{m^{2}\ell^{2}\left[b^{2}\frac{\sin^{2}{(\beta \tau)}}{\beta^{2}}\right]^
{\frac{2n (n + 1)}{n^{2} + n + 1}}} - \frac{\beta^{2}(n + 1)^{2}}{(n^{2} + n +1)}, 
\end{equation}
\begin{equation}
\label{eq42}
8\pi \rho_{p} = \frac{8n(n + 1)(n + 2)b^{2}\cos^{2}{(\beta \tau)}}{(n^{2} + n + 1)^{2}
\left[b^{2}\frac{\sin^{2}{(\beta \tau)}}{\beta^{2}}\right]^{2}}
+ \frac{4n(n + 1)\beta^{2}}{(n^{2} + n + 1)\left[b^{2}\frac{\sin^{2}{(\beta \tau)}}{\beta^{2}}
\right]} + \frac{\beta^{2}(n + 1)^{2}}{(n^{2} + n +1)}, 
\end{equation}
\begin{equation}
\label{eq43}
\theta = \frac{2(2n + 1)(n + 1)}{N(n^{2} + n + 1)} \frac{\cos{(\beta \tau)}}{\frac{\sin^{2}
{(\beta \tau)}}{\beta^{2}}},
\end{equation}
\begin{equation}
\label{eq44}
\sigma  = \frac{2(n^{2} - 1)}{\sqrt{3} N (n^{2} + n + 1)} \frac{\cos{(\beta \tau)}}
{\frac{\sin^{2}{(\beta \tau)}}{\beta^{2}}},
\end{equation}
where $b = - N, ~ N > 0$.
In absence of bulk viscosity i.e. when $\beta \to 0$ then the physical and kinematic 
quantities for the model  (\ref{eq34}) are given by 
\begin{equation}
\label{eq45}
8\pi \rho = \frac{4n(n + 2)(n + 1)^{2}}{b^{2}(n^{2} + n + 1)^{2}} \frac{1}{\tau^{4}} - 
\frac{\alpha^{2}}{m^{2}\ell^{2}}\frac{1}{[b^{4} \tau^{4}]^{\frac{n(n + 1)}{n^{2} + n + 1}}}, 
\end{equation}
\begin{equation}
\label{eq46}
8\pi \lambda = \frac{4n(n + 1)(n^{2} + n -2)}{b^{2}(n^{2} + n + 1)^{2}\tau^{4}} - \frac{\alpha^{2}}
{m^{2}\ell^{2}[b^{4} \tau^{4}]^{\frac{n(n + 1)}{n^{2} + n + 1}}},  
\end{equation}
\begin{equation}
\label{eq47}
8\pi \rho_{p} = \frac{8n(n + 1)(n + 2)}{b^{2}(n^{2} + n + 1)^{2}} \frac{1}{\tau^{4}},  
\end{equation}
\begin{equation}
\label{eq48}
\theta =  \frac{2(n^{2} - 1)}{N(n^{2} + n + 1)} \frac{1}{\tau^{2}},  
\end{equation}
\begin{equation}
\label{eq49}
\sigma =  \frac{2(n^{2} - 1)}{N\sqrt{3}(n^{2} + n + 1)} \frac{1}{\tau^{2}}.  
\end{equation}

In the absence of bulk viscosity, the energy conditions $\rho \geq 0$ and $\rho_{p} \geq 0$ are 
satisfied for the model (\ref{eq34}). The condition $\rho \geq 0$ leads to 
$$\tau^{\frac{4}{n^{2} + n + 1}} \leq \frac{4n(n + 1)^{2}(n + 2)m^{2} \ell^{2} b^{2(n^{2} + n -1)}}
{\alpha^{2}(n^{2} + n + 1)^{2}}.$$
The string tension density $\lambda \geq 0$ if 
$$\tau^{\frac{4}{n^{2} + n + 1}} \leq \frac{4n(n + 1)(n^{2} + n - 2) m^{2} \ell^{2} b^{2(n^{2} + n -1)}}
{\alpha^{2}(n^{2} + n + 1)^{2}}.$$

The model (\ref{eq34}) in the absence of bulk viscosity starts with a big bang at time $\tau = 0$ 
and the expansion in the model decreases as time increases. Since $\lim_{\tau \to \infty} 
\frac{\sigma}{\theta} \ne 0$, the model does not approach isotropy for large value of $\tau$. 
However the model isotropizes for $n = -1$ and $n = 1$. There is a real physical 
singularity in the model (\ref{eq34}) at $\tau = 0$.  

\section*{Acknowledgments} 
We  would like to thank the Inter-University Centre for Astronomy and Astrophysics
(IUCAA), Pune, India for providing facility and support where this work was carried out. 
We also thank the anonymous referees for the fruitful comments.

\end{document}